\documentclass[twocolumn,preprintnumbers,superscriptaddress]{revtex4}%
\usepackage{amssymb}
\usepackage{amsmath}
\usepackage{graphicx}
\usepackage{epstopdf}
\usepackage{dcolumn}
\usepackage{bm}
\usepackage{amsfonts}
\usepackage{xcolor}
\usepackage{ifthen}
\usepackage{soul}
\usepackage[T1]{fontenc}
\usepackage{upquote}%
\providecommand{\U}[1]{\protect\rule{.1in}{.1in}}
\providecommand{\U}[1]{\protect\rule{.1in}{.1in}}
\def\showeb{1}
\newcommand{\eb}[1]{\ifthenelse{\showeb=1}{\textcolor{cyan}{[[#1]]}}{}}
\def\showcm{1}
\newcommand{\cm}[1]{\ifthenelse{\showcm=1}{\textcolor{red}{[[#1]]}}{}}
\def\showshg{1}
\newcommand{\shg}[1]{\ifthenelse{\showshg=1}{\textcolor{blue}{[[#1]]}}{}}
\begin{document}
\title{Frequency mixing in a ferrimagnetic sphere resonator}
\author{Cijy Mathai}
\affiliation{Andrew and Erna Viterbi Department of Electrical Engineering, Technion, Haifa
32000 Israel}
\author{Sergei Masis}
\affiliation{Andrew and Erna Viterbi Department of Electrical Engineering, Technion, Haifa
32000 Israel}
\author{Oleg Shtempluck}
\affiliation{Andrew and Erna Viterbi Department of Electrical Engineering, Technion, Haifa
32000 Israel}
\author{Shay Hacohen-Gourgy}
\affiliation{Physics Department, Technion, Haifa 32000 Israel}
\author{Eyal Buks}
\affiliation{Andrew and Erna Viterbi Department of Electrical Engineering, Technion, Haifa
32000 Israel}
\date{\today }
\date{\today }
\date{\today }
\date{\today }

\begin{abstract}
Frequency mixing in ferrimagnetic resonators based on yttrium and calcium
vanadium iron garnets (YIG and CVBIG) is employed for studying their nonlinear
interactions. The ferrimagnetic Kittel mode is driven by applying a pump tone
at a frequency close to resonance. We explore two nonlinear frequency mixing
configurations. In the first one, mixing between a transverse pump tone and an
added longitudinal weak signal is explored, and the experimental results are
compared with the predictions of the Landau-Zener-Stuckelberg model. In the
second one, intermodulation measurements are employed
by mixing pump and signal tones both in the
transverse direction for studying a bifurcation between a stable spiral and a
stable node attractors. Our results are applicable for developing sensitive
signal receivers with high gain for both the radio frequency and the microwave bands.

\end{abstract}
\maketitle





\section{Introduction}

The physics of magnons in ferromagnetic resonators
\cite{Hill_S227,Lecraw_1311, Kumar_435802} has been extensively studied in the
backdrop of Bose-Einstein condensation \cite{Demokritov_430}, optomagnonics
\cite{Zhang_123605, Osada_223601, Stancil_Spin}, and spintronics
\cite{Kajiwara_262}. Owing to the high magnon life time of the order of a few
microseconds, such ferromagnetic insulators have become the natural choice of
microwave (MW) resonators in synthesizers \cite{Ryte_434}, narrow band filters
\cite{Tsai_3568}, and parametric amplifiers \cite{Kotzebue_773}. Exploring the
nonlinearity associated with such systems is gaining attention. A variety of
magnon \textit{nonlinear} dynamical effects have been studied in the context
of auto-oscillations \cite{Rezende_893}, optical cooling \cite{Sharma_087205}, frequency mixing \cite{Jepsen_2627,Morgenthaler_S157}
and bistability \cite{Wang_057202,Wang_224410,Hyde_174423,Suhl_209,Wiese_119}.
Applications of nonlinearity for quantum data processing have been explored in
\cite{Elyasi_1910_11130,Zhang_023021}. Nonlinear interactions between the
electromagnetic (EM) MW field coherent photons and these resonators can be
significantly enhanced with relatively low power around the resonance
frequency of the oscillator. Studying such nonlinear interactions is important
due to the realization of hybrid quantum systems for quantum memory and
optical transducer related applications \cite{Zhang_156401, Tabuchi_083603,
Lachance_070101,Lachance_1910_09096,Tabuchi_729,Kusminskiy_1911_11104}.

\begin{figure}
	[ptb]
	\begin{center}
		\includegraphics[width=7.5cm,keepaspectratio]%
		{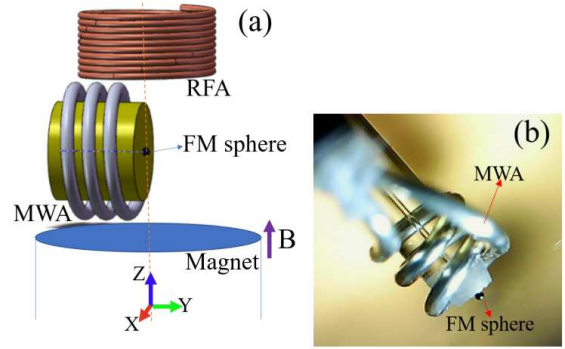}%
		\caption{(a) A schematic of the experimental setup used for studying the ferrimagnetic (FM) sphere. Longitudinal and transverse driving are applied using a radio frequency antenna (RFA) and a microwave antenna (MWA), respectively. (b) Real image of the DUT.}%
		\label{DUT}%
	\end{center}
\end{figure}

Here we study the nonlinear frequency mixing process in these ferromagnetic
resonators based on two configurations. In the first configuration we study
frequency mixing of transverse and longitudinal driving tones that are
simultaneously applied to the magnon resonator. The signal tone is in the
radio frequency (RF) band, and it is applied in the longitudinal direction,
parallel to the external static magnetic field. This process can be employed
for frequency conversion between the RF and the MW bands. Here we find that
the measured response can be well described using the Landau-Zener-Stuckelberg
model \cite{Berns_150502, Shevchenko_1}. In the second configuration, Kerr
nonlinearity that is induced by magnetic anisotropy is studied by
intermodulation measurements. This is done by simultaneously applying in the
transverse direction an intense pump and a weak signal tones both having
frequencies close to resonance. The observed intermodulation frequency
conversion reveals a bifurcation between a stable spiral and a stable node
\cite{Yurke_53}. These nonlinear effects may find applications in signal
sensing, parametric amplification and other related applications.

The spherical resonators under test are made of yttrium iron garnet (YIG) and
calcium vanadium bismuth iron garnet (CVBIG) with a radius of $R_{\mathrm{s}%
}=125%
\operatorname{\mu m}%
$. 
They host magnonic excitations with relatively low damping and large spin
densities. These spheres are anisotropic ferrimagnetic crystals with strong
Faraday rotation angles and high refractive index as compared to other iron
garnets. A schematic image of our device under test (DUT) is shown in Fig.
\ref{DUT}. The ferrimagnetic sphere is held by vacuum through a ferrule. A
fixed magnet is employed for fully magnetizing the sphere. A loop antenna
(coil) is used to apply a transverse (longitudinal) driving in the MW (RF)
band. All measurements are performed at room temperature.

\section{Landau-Zener-Stuckelberg interferometry}

Landau-Zener-Stuckelberg interferometry is based on a mixing process between
transverse and longitudinal driving frequencies that are simultaneously
applied to a resonator \cite{Berns_150502, Shevchenko_1}. The polarization
vector $\mathbf{P}$ evolves in time $t$ according to the Bloch-Landau-Lifshitz
equation $\mathrm{d}\mathbf{P/}\mathrm{d}t=\mathbf{P}\times\mathbf{\Omega
}+\mathbf{\Gamma}$, where $\mathbf{\Omega}=\gamma_{\mathrm{e}}\mathbf{B}$ is the rotation vector,
with $\mathbf{B}$ being the externally applied magnetic induction and $\gamma
_{\mathrm{e}}=28 %
\operatorname{GHz}%
\operatorname{T}%
^{-1}$ being the gyromagnetic ratio, and the vector
$\mathbf{\Gamma}=-\Gamma_{2}P_{x}\mathbf{\hat{x}}-\Gamma_{2}P_{y}%
\mathbf{\hat{y}}-\Gamma_{1}\left(  P_{z}-P_{z,\mathrm{s}}\right)
\mathbf{\hat{z}}$ represents the contribution of damping, with $\Gamma_{1}%
=1/T_{1}$ and $\Gamma_{2}=1/T_{2}$ being the longitudinal and transverse
relaxation rates, respectively, and $P_{z,\mathrm{s}}$ being the steady state
polarization. Consider the case where $\mathbf{\Omega}\left(  t\right)
=\omega_{1}\left(  \cos\left(  \omega t\right)  \mathbf{\hat{x}}+\sin\left(
\omega t\right)  \mathbf{\hat{y}}\right)  +\omega_{0}\mathbf{\hat{z}}$. Here
$\omega_{1}$ and $\omega$ are both real constants, and $\omega_{0}$ oscillates
in time according to $\omega_{0}=\omega_{\mathrm{c}}+\omega_{\mathrm{b}}%
\sin\left(  \omega_{\mathrm{m}}t\right)  $, where $\omega_{\mathrm{c}}$,
$\omega_{\mathrm{b}}$ and $\omega_{\mathrm{m}}$ are all real constants.
Nonlinearity of the Bloch-Landau-Lifshitz equation gives rise to frequency
mixing between the transverse driving at angular frequency $\omega$ and the
longitudinal driving at angular frequency $\omega_{\mathrm{m}}$. The resonance
condition of the $l$'th order frequency mixing process reads $\omega
+l\omega_{\mathrm{m}}=\omega_{\mathrm{c}}$, where $l$ is an integer. The
complex amplitude $P_{\mathrm{+}}$ (in a rotating frame) of the corresponding
$l$'th side band is given by (see appendix D of Ref. \cite{Buks_033807})%
\begin{equation}
P_{\mathrm{+}}=\frac{\frac{i\omega_{1}\zeta}{\Gamma_{2}^{2}}\left(
1+\frac{i\omega_{\mathrm{d}}}{\Gamma_{2}}\right)  P_{z,\mathrm{s}}}{\left(
1+\frac{\omega_{\mathrm{d}}^{2}}{\Gamma_{2}^{2}}\right)  \frac{\Gamma_{1}%
}{\Gamma_{2}}+\left(  \frac{\omega_{1}\zeta}{\Gamma_{2}}\right)  ^{2}}\ ,
\label{P_+}%
\end{equation}
where $\zeta=J_{l}\left(  \omega_{\mathrm{b}}/\omega_{\mathrm{m}}\right)  $,
$J_{l}$ is the $l^{\prime}$th Bessel function of the first kind, and the detuning angular frequency $\omega_{\mathrm{d}}$ is given by
$\omega_{\mathrm{d}}=\omega+l\omega_{\mathrm{m}}-\omega_{\mathrm{c}}$.

\begin{figure}[tbh]
\centering  \includegraphics[width=7.5cm,keepaspectratio]{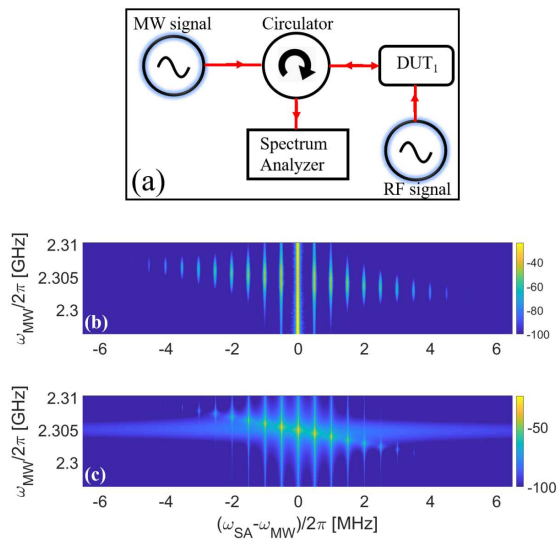}
\caption{Landau-Zener-Stuckelberg interferometry. Experimental (b) and theoretical (c) spectral response obtained by the frequency mixing of
transverse and longitudinal driving signals applied simultaneously to the
magnon resonator. The theoretical color-coded plot (c) is derived using Eq.
(\ref{P_+}), using the following parameters' values $\omega
_{\mathrm{m}}/(2\pi)=0.5 \operatorname{MHz}$, $\omega_{1}/(2\pi)=0.5
\operatorname{MHz}$, $\omega_{\mathrm{b}}/(2\pi)=0.5 \operatorname{MHz}$,
$\Gamma_{1}/(2\pi)=1.0 \operatorname{MHz}$ and $\Gamma_{2}/(2\pi)=2.0
\operatorname{MHz}$.}%
\label{Fig_LZ_setup}%
\end{figure}

The schematic of the experimental setup employed to
explore this frequency mixing process is shown
in Fig. \ref{Fig_LZ_setup} (a). The device under test (DUT$_{1}$) comprises of
the ferrimagnetic resonator coupled to both the MW loop antenna and the RF
coil. The Kittel mode frequency is tuned by the static magnetic field to the
value $2.305\operatorname{GHz}$. The sphere is simultaneously driven by a pump
with a power of 0 dBm that is applied to the MW loop antenna and an RF signal
with a frequency of $0.5\operatorname{MHz}$ that is applied to the RF coil.
Spectrum analyzer measurements of the signal reflected from the MW loop
antenna are shown in Fig. \ref{Fig_LZ_setup} (b) as a function of the spectrum
analyzer angular frequency $\omega_{\mathrm{SA}}$ and the driving MW angular
frequency $\omega_{\mathrm{MW}}$ that is injected into the loop antenna. The
theoretical prediction that is derived using Eq. (\ref{P_+}) is presented by
Fig. \ref{Fig_LZ_setup} (c). The values of parameters that are used for the
calculation are listed in the caption of Fig. \ref{Fig_LZ_setup}. The
comparison between the measured [see Fig. \ref{Fig_LZ_setup} (b)] and
calculated [see Fig. \ref{Fig_LZ_setup} (c)] response yields a good agreement.

\section{Anisotropy-induced Kerr nonlinearity}

\begin{figure}[tbh]
\includegraphics[width=7.5cm,keepaspectratio]{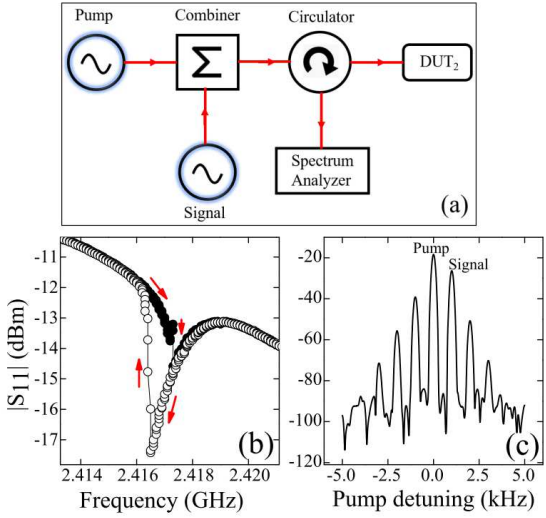}
\caption{Intermodulation. (a) An intense pump and a relatively weak signal are
simultaneously injected into the MW loop antenna, and the reflected signal is
measured using a spectrum analyzer. (b) Experimentally obtained hysteretic
resonance curve (with no signal) showing bistability corresponding to the
forward and backward microwave frequency sweep directions. (c) Spectrum
analyzer measurement of the reflected signal intensity as a function of the
detuning frequency with respect to the pump frequency. The idler peaks are
generated as a result of the nonlinear pump-signal mixing.}%
\label{IMD_bistability}%
\end{figure}

The experimental setup used for intermodulation measurements is shown in Fig.
\ref{IMD_bistability} (a). Here the device under test (DUT$_{2}$) is the same
as that shown in Fig. \ref{DUT}, where the RF antenna (RFA) is removed from
the setup. The nonlinearity gives rise to bistability, which in turn yields a
hysteretic resonance curve, which is obtained via the forward and backward
sweeping directions [see Fig. \ref{IMD_bistability} (b)]. The measured response becomes
bistable when the input pump power $P_{\mathrm{p}}$ is of the order of $%
\operatorname{mW}%
$. The subsequent idler tones generated due to the nonlinear frequency mixing
of pump and signal tones in the ferrimagnetic resonator are shown in Fig.
\ref{IMD_bistability}(c).

The technique of Bosonization can be applied to model the nonlinearity in
ferrimagnetic sphere resonators \cite{Zhang_987511}. In this approach, the
Hamiltonian $\mathcal{H}_{\mathrm{M}}$ is expressed in the form $\hbar
^{-1}\mathcal{H}_{\mathrm{M}}=\omega_{\mathrm{c}}N_{\mathrm{M}}+K_{\mathrm{M}%
}N_{\mathrm{M}}^{2}+Q_{\mathrm{M}}N_{\mathrm{M}}^{4}+\cdots$, where
$\omega_{\mathrm{c}}=\mu_{0}\gamma_{\mathrm{e}}H$ is the angular frequency of
the Kittel mode \cite{Fletcher_687,Stancil_Spin}, $\mu_{0}=4\pi\times10^{-7}%
\operatorname{N}%
\operatorname{A}%
^{-2}$ is the permeability of free space, $H$ is the externally applied
uniform magnetic field (which is assumed to be parallel to the $\mathbf{\hat
{z}}$ axis), $N_{\mathrm{M}}$ is a number operator, $K_{\mathrm{M}}$ is the
so-called Kerr frequency, and $Q_{\mathrm{M}}$ is the coefficient of quartic
nonlinearity. When nonlinearity is taken into account to lowest nonvanishing
order only, i.e. when the quartic and all higher order terms are disregarded,
the response can be described using the Duffing-Kerr model. This model
predicts that the response of the system to an externally applied
monochromatic driving can become bistable.

In general, the number of magnons $\left\langle N_{\mathrm{M}%
}\right\rangle $ in a resonantly driven sphere having total linear damping
rate $\gamma_{\mathrm{c}}$ with pump power $P_{\mathrm{p}}$ is given for the
case of critical coupling by $\left\langle N_{\mathrm{M}}\right\rangle \simeq
P_{\mathrm{p}}/\left(  \hbar\omega_{\mathrm{c}}\gamma_{\mathrm{c}}\right)  $.
On the other hand, the expected number of magnons $\left\langle N_{\mathrm{M}%
}\right\rangle $ at the onset of Duffing-Kerr bistability is $\simeq
\gamma_{\mathrm{c}}/\left\vert K_{\mathrm{M}}\right\vert $ [see Eq. (42) of
Ref. \cite{Yurke_5054} and note that, for simplicity, cubic nonlinear damping
is disregarded]. Thus, from the measured values of the linear damping rate
$\gamma_{\mathrm{c}}/\left(  2\pi\right)  \simeq1%
\operatorname{MHz}%
$ and $P_{\mathrm{p}}\simeq1%
\operatorname{mW}%
$, at the bistability onset point one obtains $K_{\mathrm{M}}/\left(
2\pi\right)  \simeq-2\times10^{-9}%
\operatorname{Hz}%
$ (the minus signs indicates that the Kerr nonlinearity gives rise to
softening). Note, however, that the above estimate, which is based on the
Duffing-Kerr model, is valid provided that the quartic and all higher order
terms can be disregarded near (and below) the bistability onset. For the
quartic term this condition can be expressed as $\left\vert Q_{\mathrm{M}}\right\vert \ll\left\vert
K_{\mathrm{M}}\right\vert ^{3}/\gamma_{\mathrm{c}}^{2}$.

The values of $K_{\mathrm{M}}$ and $Q_{\mathrm{M}}$ are estimated below for
the case where nonlinearity originates from magnetic anisotropy. The
Stoner--Wohlfarth energy $E_{\mathrm{M}}$ is expressed as a function of the
magnetization vector $\mathbf{M}=M\mathbf{\hat{u}}_{\mathrm{M}}$, and the
first-order $K_{\mathrm{c1}}$ and second-order $K_{\mathrm{c2}}$\ anisotropy
constants as \cite{Blunde_Mag}%
\begin{equation}
\frac{E_{\mathrm{M}}}{V_{\mathrm{s}}}=-\mu_{0}\mathbf{M}\cdot\mathbf{H}%
+K_{\mathrm{c1}}\sin^{2}\phi+K_{\mathrm{c2}}\sin^{4}\phi\;, \label{E_M}%
\end{equation}
where $V_{\mathrm{s}}=4\pi R_{\mathrm{s}}^{3}/3$ is the volume of the sphere
having radius $R_{\mathrm{s}}$, and $\phi$ is the angle between $\mathbf{\hat
{u}}_{\mathrm{M}}$ and the unit vector $\mathbf{\hat{u}}_{\mathrm{A}}$
parallel to the easy axis. It is assumed that the sphere is fully magnetized,
i.e. $\left\vert \mathbf{M}\right\vert \simeq M_{\mathrm{s}}$, where
$M_{\mathrm{s}}$ is the saturation magnetization. In terms of the
dimensionless angular momentum vector $\mathbf{\Sigma}=-2\mathbf{M}%
V_{\mathrm{s}}/\left(  \hbar\gamma_{\mathrm{e}}\right)  \equiv\left(
\Sigma_{x},\Sigma_{y},\Sigma_{z}\right)  $ Eq. (\ref{E_M}) is rewritten as
$E_{\mathrm{M}}-\hbar\omega_{\mathrm{K1}}\left(  1+K_{\mathrm{c2}%
}/K_{\mathrm{c1}}\right)  =\mathcal{H}_{\mathrm{M}}$, where%
\begin{align}
\hbar^{-1}\mathcal{H}_{\mathrm{M}}  &  =\frac{\omega_{\mathrm{c}}\Sigma_{z}%
}{2}+\left(  1+\frac{2K_{\mathrm{c2}}}{K_{\mathrm{c1}}}\right)  \frac
{K_{\mathrm{M}}\left(  \mathbf{\Sigma}\cdot\mathbf{\hat{u}}_{\mathrm{A}%
}\right)  ^{2}}{4}\nonumber\\
&  +\frac{K_{\mathrm{c2}}}{K_{\mathrm{c1}}}\frac{K_{\mathrm{M}}^{2}\left(
\mathbf{\Sigma}\cdot\mathbf{\hat{u}}_{\mathrm{A}}\right)  ^{4}}{16\omega
_{\mathrm{K1}}}\;,\nonumber\\
&  \label{H_M}%
\end{align}
$\omega_{\mathrm{K1}}=\hbar^{-1}V_{\mathrm{s}}K_{\mathrm{c1}}$ and
$K_{\mathrm{M}}=\hbar\gamma_{\mathrm{e}}^{2}K_{\mathrm{c1}}/\left(
V_{\mathrm{s}}M_{\mathrm{s}}^{2}\right)  $ is the Kerr frequency
\cite{Wang_224410}.

In the Holstein-Primakoff transformation \cite{Holstein_1098}, the operators
$\Sigma_{\pm}=\Sigma_{x}\pm i\Sigma_{y}$ and $\Sigma_{z}$ are expressed as
$\Sigma_{+}=B^{\dag}\left(  N_{\mathrm{s}}-N_{\mathrm{M}}^{\prime}\right)
^{1/2}$, $\Sigma_{-}=\left(  N_{\mathrm{s}}-N_{\mathrm{M}}^{\prime}\right)
^{1/2}B$ and $\Sigma_{z}=-N_{\mathrm{s}}+2N_{\mathrm{M}}^{\prime}$, where
$N_{\mathrm{s}}$ is the total number of spins, and where $N_{\mathrm{M}%
}^{\prime}=B^{\dag}B$ is a number operator. If the operator $B$ satisfies the
Bosonic commutation relation $\left[  B,B^{\dag}\right]  =1$ then the
following holds $\left[  \Sigma_{z},\Sigma_{+}\right]  =2\Sigma_{+}$, $\left[
\Sigma_{z},\Sigma_{-}\right]  =-2\Sigma_{-}$ and $\left[  \Sigma_{+}%
,\Sigma_{-}\right]  =\Sigma_{z}$. The approximation $\left(  N_{\mathrm{s}%
}-N_{\mathrm{M}}^{\prime}\right)  ^{1/2}\simeq N_{\mathrm{s}}^{1/2}$ leads to
$\mathbf{\Sigma}\cdot\mathbf{\hat{u}}_{\mathrm{A}}=N_{\mathrm{s}}^{1/2}\left(
B^{\dag}u_{\mathrm{A}+}+Bu_{\mathrm{A}-}\right)  +2N_{\mathrm{M}}%
u_{\mathrm{A}z}$, where $u_{\mathrm{A}\pm}=\left[  \left(  \mathbf{\hat{u}%
}_{\mathrm{A}}\cdot\mathbf{\hat{x}}\right)  \mp i\left(  \mathbf{\hat{u}%
}_{\mathrm{A}}\cdot\mathbf{\hat{y}}\right)  \right]  /2$, $u_{\mathrm{A}%
z}=\mathbf{\hat{u}}_{\mathrm{A}}\cdot\mathbf{\hat{z}}$, and the magnon number
operator $N_{\mathrm{M}}$ is defined by $N_{\mathrm{M}}=N_{\mathrm{M}}%
^{\prime}-N_{\mathrm{s}}/2$. This approximation is valid near the bistability
onset provided that $\gamma_{\mathrm{c}}/\left(  \left\vert K_{\mathrm{M}%
}\right\vert N_{\mathrm{s}}\right)  \ll1$. For YIG, the spin density is
$\rho_{\mathrm{s}}=4.2\times10^{21}%
\operatorname{cm}%
^{-3}$, thus for a sphere of radius $R_{\mathrm{s}}=125%
\operatorname{\mu m}%
$ the number of spins is $N_{\mathrm{s}}=V_{\mathrm{s}}\rho_{\mathrm{s}%
}=3.\,4\times10^{16}$, hence for the current experiment $\gamma_{\mathrm{c}%
}/\left(  \left\vert K_{\mathrm{M}}\right\vert N_{\mathrm{s}}\right)
\simeq10^{-1}$. This estimate suggests that inaccuracy originating from this
approximation may be significant for the current experiment near and above the
bistability threshold.

Second-order\ anisotropy gives rise to a quartic nonlinear term in the
Hamiltonian (\ref{H_M}) with a coefficient $Q_{\mathrm{M}}\simeq\left(
K_{\mathrm{c2}}/K_{\mathrm{c1}}\right)  \left(  K_{\mathrm{M}}^{2}%
/\omega_{\mathrm{K1}}\right)  $ (the exact value depends on the angle $\phi$
between the magnetization vector and the easy axis). Near or below the
bistability onset the quartic term can be safely disregarded provided that
$\left(  K_{\mathrm{c2}}/K_{\mathrm{c1}}\right)  \left(  \gamma_{\mathrm{c}%
}^{2}/\left(  \omega_{\mathrm{K1}}\left\vert K_{\mathrm{M}}\right\vert
\right)  \right)  \ll1$. When this condition is satisfied the Hamiltonian
(\ref{H_M}) for the case where $\mathbf{\hat{u}}_{\mathrm{A}}$ is parallel to
$\mathbf{\hat{z}}$ (i.e. $u_{\mathrm{A}z}=1$ and $u_{\mathrm{A}+}%
=u_{\mathrm{A}-}=0$) approximately becomes%
\begin{equation}
\hbar^{-1}\mathcal{H}_{\mathrm{M}}=\omega_{\mathrm{c}}N_{\mathrm{M}%
}+K_{\mathrm{M}}N_{\mathrm{M}}^{2}\;. \label{H Kerr}%
\end{equation}
The term proportional to $K_{\mathrm{M}}$ represents the anisotropy-induced
Kerr nonlinearity.

For YIG $M_{\mathrm{s}}=140%
\operatorname{kA}%
/%
\operatorname{m}%
$, $K_{\mathrm{c1}}=-610%
\operatorname{J}%
/%
\operatorname{m}%
^{3}$ at $297%
\operatorname{K}%
$ (room temperature), hence for a sphere of radius $R_{\mathrm{s}}=125%
\operatorname{\mu m}%
$ the expected value of the Kerr coefficient is given by $K_{\mathrm{M}%
}/\left(  2\pi\right)  =-2.0\times10^{-9}%
\operatorname{Hz}%
$. This value well agrees with the above estimation of $K_{\mathrm{M}}/\left(
2\pi\right)  $ based on the measured input power at the bistability onset. For
YIG $K_{\mathrm{c2}}/K_{\mathrm{c1}}=4.8\times10^{-2}$ ($K_{\mathrm{c2}%
}/K_{\mathrm{c1}}=4.3\times10^{-2}$) at a temperature of $T=4.2%
\operatorname{K}%
$ ($T=294%
\operatorname{K}%
$) \cite{Stancil_Spin}. Based on these values one finds that for the sphere
resonators used in the current experiment $\left(  K_{\mathrm{c2}%
}/K_{\mathrm{c1}}\right)  \left(  \gamma_{\mathrm{c}}^{2}/\left(
\omega_{\mathrm{K1}}\left\vert K_{\mathrm{M}}\right\vert \right)  \right)
\simeq10^{-6}$, hence the second-order\ anisotropy term (proportional to
$K_{\mathrm{c2}}$) in Eq. (\ref{H_M}) can be safely disregarded in the
vicinity of the bistability onset.

\section{Stable spiral and stable node}

To explore the regime of weak nonlinear response, consider a resonator
being driven by a monochromatic pump tone having
amplitude $b_{\mathrm{c}}$ and angular frequency $\omega_{\mathrm{p}}$. The
time evolution in a frame rotating at the pump driving frequency is assumed to
have the form
\begin{equation}
\frac{\mathrm{d}C_{\mathrm{c}}}{\mathrm{d}t}+\Theta_{\mathrm{c}}%
=F_{\mathrm{c}}\;,\label{dC/dt}%
\end{equation}
where the operator $C_{\mathrm{c}}$\ is related to the resonator's
annihilation operator $A_{\mathrm{c}}$ by $C_{\mathrm{c}}=A_{\mathrm{c}%
}e^{i\omega_{\mathrm{p}}t}$, the term $\Theta_{\mathrm{c}}=\Theta_{\mathrm{c}%
}\left(  C_{\mathrm{c}},C_{\mathrm{c}}^{\dag}\right)  $, which is expressed as
a function of both $C_{\mathrm{c}}$ and $C_{\mathrm{c}}^{\dag}$, is assumed to
be time independent, and $F_{\mathrm{c}}$ is a noise term having a vanishing
expectation value. The complex number $B_{\mathrm{c}}$ represents a fixed
point, for which $\Theta_{\mathrm{c}}\left(  B_{\mathrm{c}},B_{\mathrm{c}%
}^{\ast}\right)  =0$. By expressing the solution as $C_{\mathrm{c}%
}=B_{\mathrm{c}}+c_{\mathrm{c}}$ and considering the operator $c_{\mathrm{c}}$
as small, one obtains a linearized equation of motion from Eq. (\ref{dC/dt})
given by%
\begin{equation}
\frac{\mathrm{d}c_{\mathrm{c}}}{\mathrm{d}t}+W_{1}c_{\mathrm{c}}%
+W_{2}c_{\mathrm{c}}^{\dag}=F_{\mathrm{c}}\;,\label{dc_c/dt}%
\end{equation}
where $W_{1}=\partial\Theta_{\mathrm{c}}/\partial C_{\mathrm{c}}$ and
$W_{2}=\partial\Theta_{\mathrm{c}}/\partial C_{\mathrm{c}}^{\dag}$ (both
derivatives are evaluated at the fixed point $C_{\mathrm{c}}=B_{\mathrm{c}}$).

The stability properties of the fixed point depend on the eigenvalues
$\lambda_{\mathrm{c}1}$ and $\lambda_{\mathrm{c}2}$ of the $2\times2$ matrix
$W$, whose elements are given by $W_{11}=W_{22}^{\ast}=W_{1}$ and
$W_{12}=W_{21}^{\ast}=W_{2}$ [see Eq. (\ref{dc_c/dt})]. In terms of the trace
$T_{\mathrm{W}}=W_{1}+W_{1}^{\ast}$ and the determinant $D_{\mathrm{W}%
}=\left\vert W_{1}\right\vert ^{2}-\left\vert W_{2}\right\vert ^{2}$ of the
matrix $W$, the eigenvalues are given by $\lambda_{\mathrm{c}1}=T_{\mathrm{W}%
}/2+\upsilon_{\mathrm{W}}$ and $\lambda_{\mathrm{c}2}=T_{\mathrm{W}%
}/2-\upsilon_{\mathrm{W}}$, where the coefficient $\upsilon_{\mathrm{W}}$ is
given by $\upsilon_{\mathrm{W}}=\sqrt{\left(  T_{\mathrm{W}}/2\right)
^{2}-D_{\mathrm{W}}}$. Note that in the linear regime, i.e. when $W_{2}=0$,
the eigenvalues become $\lambda_{\mathrm{c}1}=W_{1}$ and $\lambda
_{\mathrm{c}2}=W_{1}^{\ast}$. For the general case, when both $\lambda
_{\mathrm{c}1}$ and $\lambda_{\mathrm{c}2}$ have a positive real part, the
fixed point is locally stable. Two types of stable fixed points can be
identified. For the so-called stable spiral, the coefficient $\upsilon
_{\mathrm{W}}$ is pure imaginary [i.e. $\left(  T_{\mathrm{W}}/2\right)
^{2}-D_{\mathrm{W}}<0$], and consequently $\lambda_{\mathrm{c}2}%
=\lambda_{\mathrm{c}1}^{\ast}$, whereas both $\lambda_{\mathrm{c}1}$ and
$\lambda_{\mathrm{c}2}$ are pure real for the so-called stable node, for which
$\upsilon_{\mathrm{W}}$ is pure real. A bifurcation between a stable spiral
and a stable node occurs when $\upsilon_{\mathrm{W}}$ vanishes.

Further insight can be gained by geometrically analyzing the dynamics near an
attractor. To that end the operators $c_{\mathrm{c}}$ and $F_{\mathrm{c}}%
$\ are treated as complex numbers. The equation of motion (\ref{dc_c/dt}) for
the complex variable $c_{\mathrm{c}}$ can be rewritten as $\mathrm{d}\bar{\xi
}/\mathrm{d}t+W^{\prime}\bar{\xi}=\bar{f}$, where $\bar{\xi}=\left(
\operatorname{Real}\left(  c_{\mathrm{c}}e^{i\phi}\right)
,\operatorname*{Imag}\left(  c_{\mathrm{c}}e^{i\phi}\right)  \right)^{\mathrm{T}}  $ and
$\bar{f}=\left(  \operatorname{Real}\left(  F_{\mathrm{c}}e^{i\phi}\right)
,\operatorname*{Imag}\left(  F_{\mathrm{c}}e^{i\phi}\right)  \right)^{\mathrm{T}}  $ are both
two-dimensional real vectors, and where the rotation angle $\phi$ is real.
Transformation into the so-called system of principle axes is obtained when
the angle $\phi$ is taken to be given by $e^{2i\phi}=W_{1}W_{2}^{\ast
}/\left\vert W_{1}W_{2}\right\vert $. For this case the \thinspace$2\times2$
real matrix $W^{\prime}$ becomes%
\begin{equation}
W^{\prime}=\left(
\begin{array}
[c]{cc}%
\cos\theta_{1} & -\sin\theta_{1}\\
\sin\theta_{1} & \cos\theta_{1}%
\end{array}
\right)  \left(
\begin{array}
[c]{cc}%
W_{+} & 0\\
0 & W_{-}%
\end{array}
\right)  \;,
\end{equation}
where $\theta_{1}=\arg\left(  W_{1}\right)  $ and where $W_{\pm}=\left\vert
W_{1}\right\vert \pm\left\vert W_{2}\right\vert $. Thus, multiplication by the
matrix $W^{\prime}$ can be interpreted for this case as a squeezing with
coefficients $W_{\pm}$ followed by a rotation by the angle $\theta_{1}$.

The flow near an attractor is governed by the eigenvectors of the $2\times2$
real matrix $W^{\prime}$. For the case where $\upsilon_{\mathrm{W}}$ is pure
real the angle $\alpha_{\mathrm{W}}$ between these eigenvectors is found to be
given by $\sin\alpha_{\mathrm{W}}=\upsilon_{\mathrm{W}}/\left\vert
W_{2}\right\vert $. Thus at the bifurcation between a stable spiral and a
stable node, i.e. when $\upsilon_{\mathrm{W}}=0$, the two eigenvectors of
$W^{\prime}$ become parallel to one another. In the opposite limit, when
$\upsilon_{\mathrm{W}}=\left\vert W_{2}\right\vert $, i.e. when $W_{1}$
becomes real, and consequently the matrix $W$ becomes Hermitian, the two
eigenvectors become orthogonal to one another (i.e. $\alpha_{\mathrm{W}}%
=\pi/2$).

The bifurcation between a stable spiral and a stable node can be observed by
measuring the intermodulation conversion gain $G_{\mathrm{I}}$ of the
resonator. This is done by injecting another input tone (in addition to the
pump tone), which is commonly referred to as the signal, at angular frequency
$\omega_{\mathrm{p}}+\omega$. The intermodulation gain is defined by
$G_{\mathrm{I}}\left(  \omega\right)  =\left\vert g_{\mathrm{I}}\left(
\omega\right)  \right\vert ^{2}$, where $g_{\mathrm{I}}\left(  \omega\right)
$ is the ratio between the output tone at angular frequency $\omega
_{\mathrm{p}}-\omega$, which is commonly referred to as the idler, and the
input signal at angular frequency $\omega_{\mathrm{p}}+\omega$. In terms of
the eigenvalues $\lambda_{\mathrm{c}1}$ and $\lambda_{\mathrm{c}2}$ the gain
$G_{\mathrm{I}}$ is given by \cite{Yurke_5054}%
\begin{equation}
G_{\mathrm{I}}=\left\vert \frac{2\gamma_{\mathrm{c}1}W_{2}}{\left(
\lambda_{\mathrm{c}1}-i\omega\right)  \left(  \lambda_{\mathrm{c}2}%
-i\omega\right)  }\right\vert ^{2}\;, \label{G_I}%
\end{equation}
where $\gamma_{\mathrm{c}1}$ is the coupling coefficient (in units of rate)
between the feedline that is used to deliver the input and output signals and
the resonator. For the case of a stable spiral, i.e. when $\lambda
_{\mathrm{c}2}=\lambda_{\mathrm{c}1}^{\ast}$, one has $\left\vert \left(
\lambda_{\mathrm{c}1}-i\omega\right)  \left(  \lambda_{\mathrm{c}2}%
-i\omega\right)  \right\vert ^{2}=\left[  \lambda^{\prime2}+\left(
\lambda^{\prime\prime}-\omega\right)  ^{2}\right]  \left[  \lambda^{\prime
2}+\left(  \lambda^{\prime\prime}+\omega\right)  ^{2}\right]  $, where
$\lambda^{\prime}=\operatorname{Re}\lambda_{\mathrm{c}1}$ and $\lambda
^{\prime\prime}=\operatorname{Im}\lambda_{\mathrm{c}1}$ (i.e. $\lambda
_{\mathrm{c}1}=\lambda^{\prime}+i\lambda^{\prime\prime}$), whereas for the
case of a stable node, i.e. when both $\lambda_{\mathrm{c}1}$ and
$\lambda_{\mathrm{c}2}$ are pure real, one has $\left\vert \left(
\lambda_{\mathrm{c}1}-i\omega\right)  \left(  \lambda_{\mathrm{c}2}%
-i\omega\right)  \right\vert ^{2}=\left(  \lambda_{\mathrm{c}1}^{2}+\omega
^{2}\right)  \left(  \lambda_{\mathrm{c}2}^{2}+\omega^{2}\right)  $.

\begin{figure}[ptb]
\begin{center}
\includegraphics[width=4in,height=2.5in,keepaspectratio]		{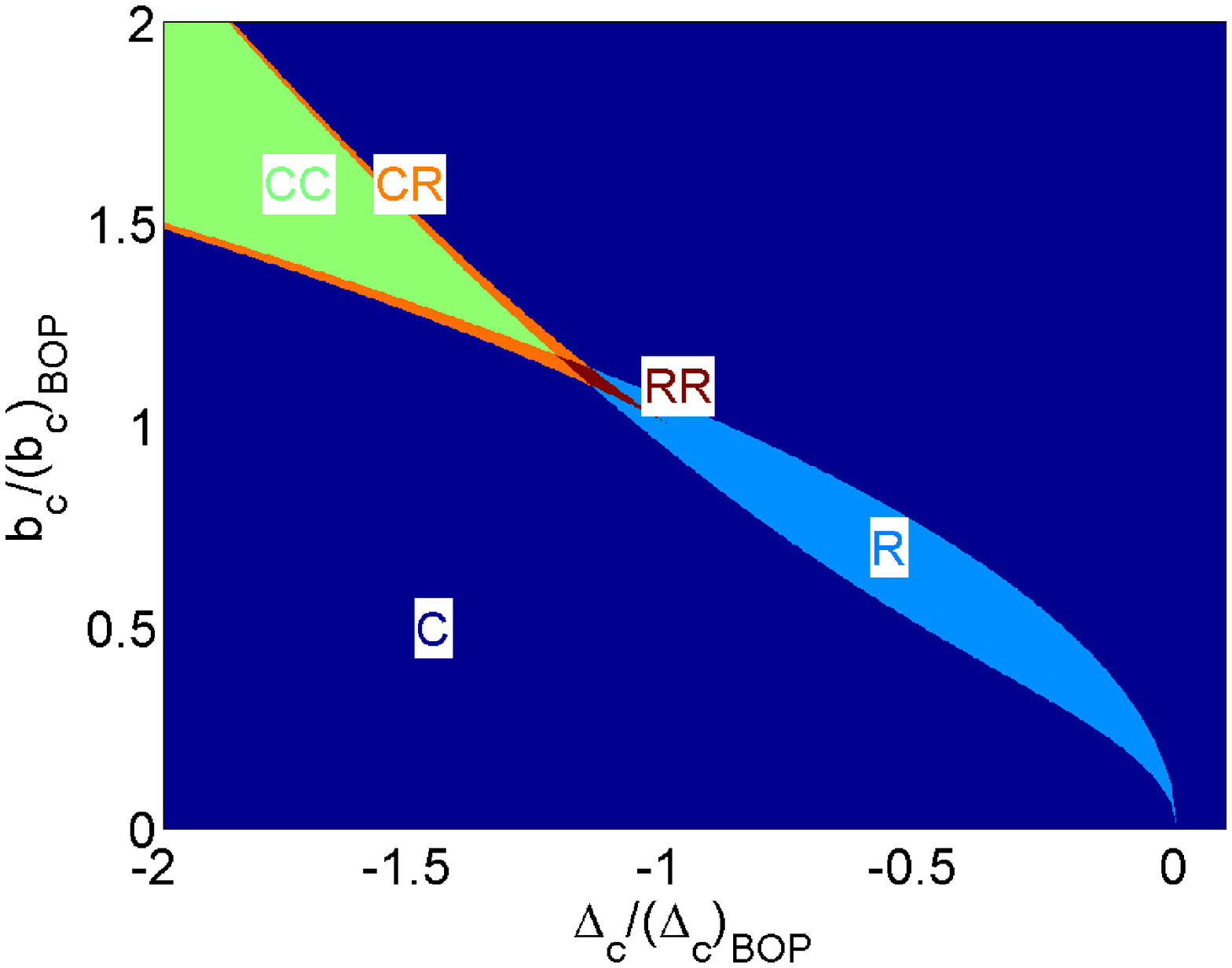}
\end{center}
\caption{Stability map of a driven Duffing-Kerr resonator. The BOP is the
point $\left(  \Delta_{\mathrm{c}}/\left(  \Delta_{\mathrm{c}}\right)
_{\mathrm{BOP}},b_{\mathrm{c}}/\left(  b_{\mathrm{c}}\right)  _{\mathrm{BOP}
}\right)  =\left(  -1,1\right)  $. In both \textquotesingle $\mathrm{C}%
$\textquotesingle $\hspace{0.1 cm}$(dark blue) and \textquotesingle$\mathrm{R}%
$\textquotesingle $\hspace{0.1 cm}$(light blue) regions, there is a single
locally stable attractor, whereas there are two in the regions
\textquotesingle$\mathrm{CC}$\textquotesingle $\hspace{0.1 cm}$(light green),
\textquotesingle$\mathrm{CR}$\textquotesingle $\hspace{0.1 cm}$(orange) and
\textquotesingle$\mathrm{RR}$\textquotesingle $\hspace{0.1 cm}$(red). The
letter \textquotesingle C\textquotesingle $\hspace{0.1 cm}$is used to label a
stable spiral, whereas the letter \textquotesingle$\mathrm{R}$%
\textquotesingle $\hspace{0.1 cm}$labels a stable node.}%
\label{FigDSM}%
\end{figure}

For the case of a resonator having Kerr nonlinearity and cubic nonlinear
damping $\Theta_{\mathrm{c}}$\ is given by $\Theta_{\mathrm{c}}=\left[
i\Delta_{\mathrm{c}}+\gamma_{\mathrm{c}}+\left(  iK_{\mathrm{c}}%
+\gamma_{\mathrm{c}3}\right)  N_{\mathrm{c}}\right]  C_{\mathrm{c}}%
+i\sqrt{2\gamma_{\mathrm{c}1}}e^{i\phi_{\mathrm{c}1}}b_{\mathrm{c}}$, where
$\Delta_{\mathrm{c}}=\omega_{\mathrm{c}}-\omega_{\mathrm{p}}$ is the driving
detuning, the total rate of linear damping is $\gamma_{\mathrm{c}}%
=\gamma_{\mathrm{c}1}+\gamma_{\mathrm{c}2}$, the rate $\gamma_{\mathrm{c}1}$
characterizes the coupling coefficient between the feedline and the resonator,
$\gamma_{\mathrm{c}2}$ is the rate of internal linear damping, $\gamma
_{\mathrm{c}3}$ is the rate of internal cubic damping, $K_{\mathrm{c}}$ is the
Kerr coefficient, $N_{\mathrm{c}}=A_{\mathrm{c}}^{\dag}A_{\mathrm{c}}$ is the
resonator number operator, and $\phi_{\mathrm{c}1}$ is a phase coefficient
characterizing the coupling between the feedline and the resonator
\cite{Yurke_5054}. The rates $W_{1}$ and $W_{2}$ are given by $W_{1}%
=i\Delta_{\mathrm{c}}+\gamma_{\mathrm{c}}+2\left(  iK_{\mathrm{c}}%
+\gamma_{\mathrm{c}3}\right)  \left\vert B_{\mathrm{c}}\right\vert ^{2}$ and
$W_{2}=\left(  iK_{\mathrm{c}}+\gamma_{\mathrm{c}3}\right)  B_{\mathrm{c}}%
^{2}$. The condition $\Theta_{\mathrm{c}}\left(  B_{\mathrm{c}},B_{\mathrm{c}%
}^{\ast}\right)  =0$ can be expressed as a cubic polynomial equation for the
number of magnons $E_{\mathrm{c}}=\left\vert B_{\mathrm{c}}\right\vert ^{2}$
given by $\left[  \left(  \Delta_{\mathrm{c}}+K_{\mathrm{c}}E_{\mathrm{c}%
}\right)  ^{2}+\left(  \gamma_{\mathrm{c}}+\gamma_{\mathrm{c}3}E_{\mathrm{c}%
}\right)  ^{2}\right]  E_{\mathrm{c}}=2\gamma_{\mathrm{c}1}\left\vert
b_{\mathrm{c}}\right\vert ^{2}$. The eigenvalues can be expressed in terms of
$E_{\mathrm{c}}$ as $\lambda_{\mathrm{c}1,2}=T_{\mathrm{W}}/2\pm
\upsilon_{\mathrm{W}}$, where $T_{\mathrm{W}}/2=\gamma_{\mathrm{c}}%
+2\gamma_{\mathrm{c}3}E_{\mathrm{c}}$ and $\upsilon_{\mathrm{W}}=\sqrt{\left(
\Delta_{-}-\Delta_{\mathrm{c}}\right)  \left(  \Delta_{+}+\Delta_{\mathrm{c}%
}\right)  }$, where $\Delta_{\pm}=\left(  \sqrt{1+\left(  \gamma_{\mathrm{c}%
3}/K_{\mathrm{c}}\right)  ^{2}}\pm2\right)  K_{\mathrm{c}}E_{\mathrm{c}}$. The
stability map of the system is shown in Fig. \ref{FigDSM}. Both driving
detuning $\Delta_{\mathrm{c}}$ and driving amplitude $b_{\mathrm{c}}$ are
normalized with the corresponding values at the bistability onset point (BOP)
$\left(  \Delta_{\mathrm{c}}\right)  _{\mathrm{BOP}}$ and $\left(
b_{\mathrm{c}}\right)  _{\mathrm{BOP}}$ [see Eqs. (46) and (47) of Ref.
\cite{Yurke_5054}]. Inside the regions \textquotesingle$\mathrm{C}%
$\textquotesingle$\hspace{0.025cm}$ and \textquotesingle$\mathrm{R}%
$\textquotesingle$\hspace{0.025cm}$ of mono-stability
(\textquotesingle$\mathrm{CC}$\textquotesingle, \textquotesingle$\mathrm{CR}%
$\textquotesingle$\hspace{0.025cm}$ and \textquotesingle$\mathrm{RR}%
$\textquotesingle$\hspace{0.025cm}$ of bistability) the resonator has one
(two) locally stable attractors. A stable spiral (node), for which
$\lambda_{\mathrm{c}2}=\lambda_{\mathrm{c}1}^{\ast}$ (both $\lambda
_{\mathrm{c}1}$ and $\lambda_{\mathrm{c}2}$ are pure real), is labeled by
\textquotesingle$\mathrm{C}$\textquotesingle$\hspace{0.025cm}$
(\textquotesingle$\mathrm{R}$\textquotesingle). 

\begin{figure}[ptb]
\begin{center}
\includegraphics[
		height=5.5in,
		width=3.4546in
		]{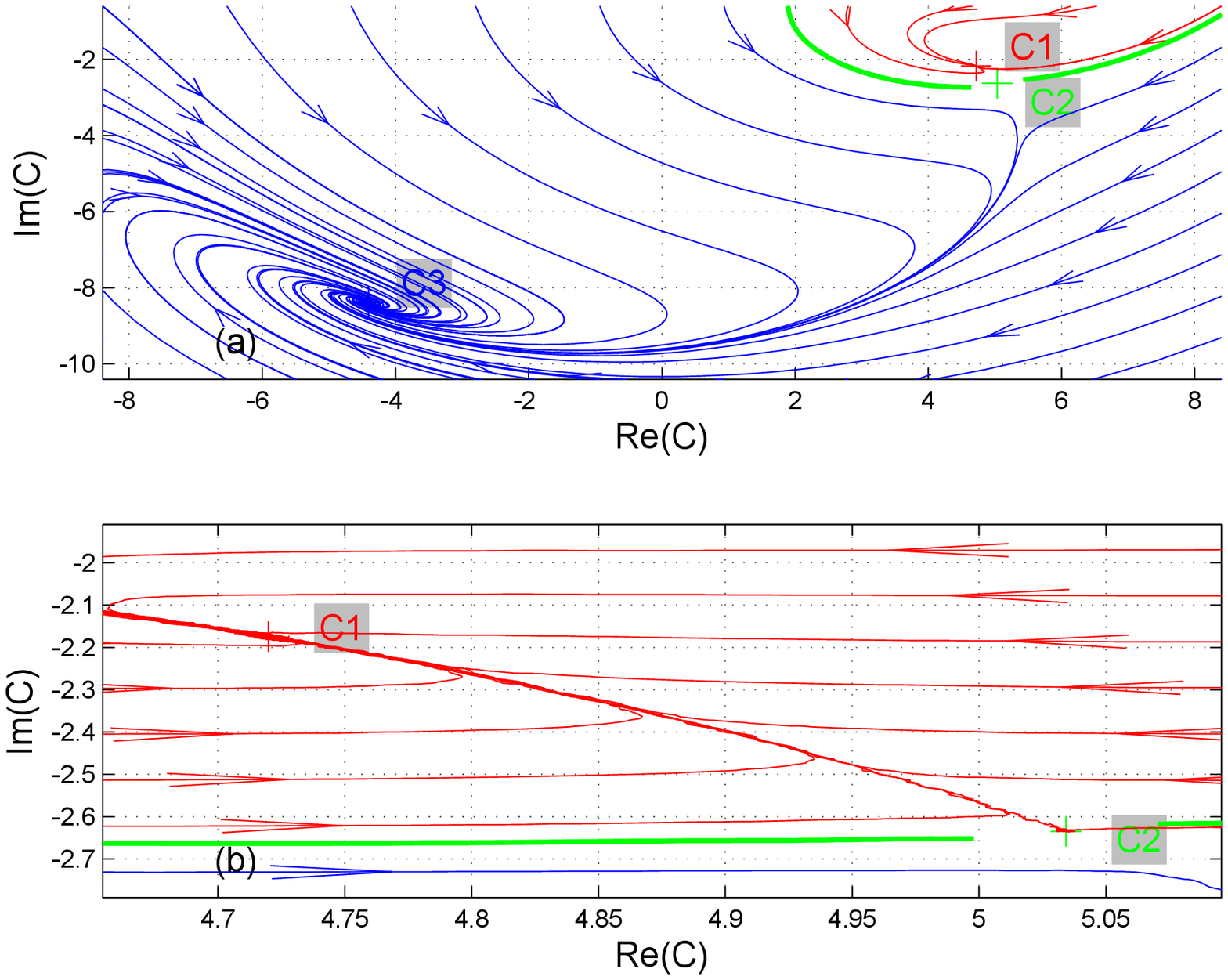}
\end{center}
\caption{Flow map of a Duffing oscillator in the region of bistability. The
point $C_{1}$ is a stable node, the point $C_{2}$ is a saddle point, and the
point $C_{3}$ is a stable spiral. A closer view of the region near $C_{1}$ and
$C_{2}$ is shown in (b). }%
\label{Fig_DuffFlow}%
\end{figure}

In the bistable region, the cubic polynomial equation has 3 real solutions for
$E_{\mathrm{c}}$. The corresponding values of the complex amplitude
$B_{\mathrm{c}}$ are labeled as $C_{1}$, $C_{2}$ and $C_{3}$. In the flow map
shown in Fig. \ref{Fig_DuffFlow}, which is obtained by numerically integrating
the equation of motion (\ref{dC/dt}) for the noiseless case $F_{\mathrm{c}}%
=0$, the point $C_{1}$ is a stable node, the point $C_{2}$ is a saddle point
and the point $C_{3}$ is a stable spiral.  The red and blue lines represent
flow toward the stable node attractor at $C_{1}$ and the stable spiral
attractor at $C_{3}$, respectively. The green line is the seperatrix, namely
the boundary between the basins of attraction of the attractors at $C_{1}$ and
$C_{3}$. A closer view of the region near $C_{1}$ and $C_{2}$ is shown in Fig.
\ref{Fig_DuffFlow}(b). 

\begin{figure}[ptb]
\begin{center}
\includegraphics[width=4in,height=2.8in,keepaspectratio]		{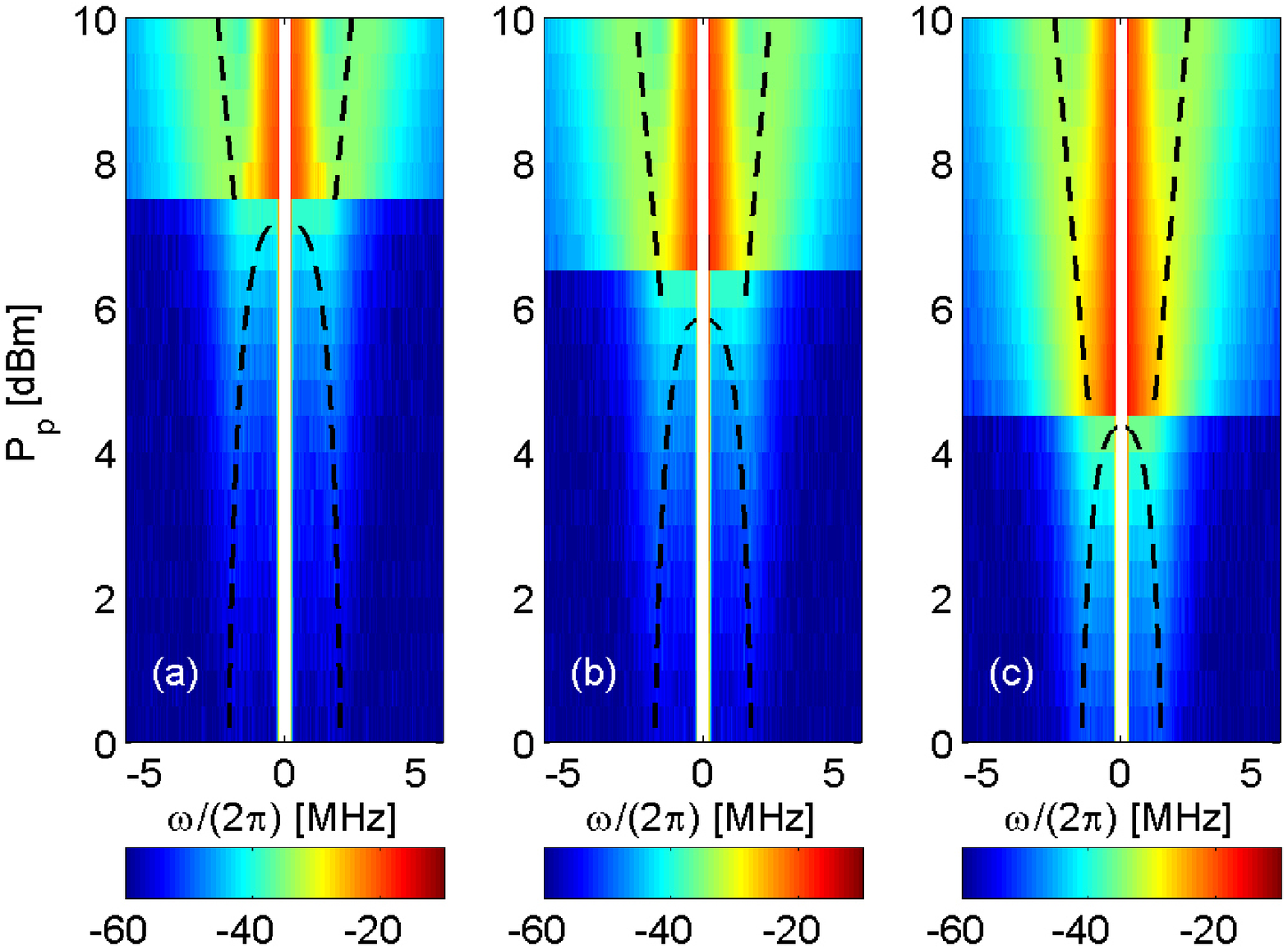}
\end{center}
\caption{Intermodulation gain $G_{\mathrm{I}}$ as a function of detuning
between the signal and pump frequencies $\omega/\left(  2\pi\right)  $ and
pump power $P_{\mathrm{p}}$ (in dBm units). The pump frequency $\omega
_{\mathrm{p}}/\left(  2\pi\right)  $ is (a) $3.8674\operatorname{GHz}$ (b)
$3.8704\operatorname{GHz}$ and (c) $3.8734\operatorname{GHz}$. The signal
power is $-15$ dBm. Note that $G_{\mathrm{I}}$ is measured with $\omega>0$
only, and the plots are generated by mirror reflection of the data around the
point $\omega=0$. Note also that for clarity the region near the pump
frequency, i.e. close to $\omega=0$, has been removed from the plot. The width
of this region, in which the intense pump peak is observed, depends on the
resolution bandwidth setting of the spectrum analyzer. The black dotted lines
indicate the calculated values of the imaginary part of the eigenvalues
$\lambda^{\prime\prime}=\operatorname{Im}\lambda_{\mathrm{c}1}$ and
$-\lambda^{\prime\prime}=\operatorname{Im}\lambda_{\mathrm{c}2}$. The pump
amplitude $b_{\mathrm{c}}$ and pump detuning $\Delta_{\mathrm{c}}$ used for
the calculation of the eigenvalues are determined from the measured value of
$P_{\mathrm{p}}=0.4$ dBm for the pump power and the value of $\Delta
_{\mathrm{c}}/\left(  2\pi\right)  =1.3\operatorname{MHz}$ for the pump
detuning at the BOP.}%
\label{FigIG}%
\end{figure}

The intermodulation conversion gain $G_{\mathrm{I}}$ induced by the Kerr
nonlinearity is measured with the ferrimagnetic resonator DUT$_{2}$ [see Fig.
\ref{IMD_bistability}(a)], and the results are compared with the theoretical
prediction given by Eq. (\ref{G_I}). In these measurements the pump frequency
$\omega_{\mathrm{p}}$ is tuned close to the resonance frequency $\omega
_{\mathrm{c}}$. The measured gain $G_{\mathrm{I}}$ is shown in the color-coded
plots in Fig. \ref{FigIG} (for three different values of the pump frequency
$\omega_{\mathrm{p}}$) as a function of the detuning between the signal and
pump frequencies $\omega/\left(  2\pi\right)  $ and the pump power
$P_{\mathrm{p}}$.

The overlaid black dotted lines in Fig. \ref{FigIG} indicate the calculated
values of the imaginary part of the eigenvalues $\lambda^{\prime\prime
}=\operatorname{Im}\lambda_{\mathrm{c}1}$ and $-\lambda^{\prime\prime
}=\operatorname{Im}\lambda_{\mathrm{c}2}$. The calculation is based on the
above-discussed Duffing-Kerr model. At the point where $\lambda^{\prime\prime
}$ vanishes, a bifurcation from stable spiral to stable node occurs. As can be
seen from comparing panels (a), (b) and (c) of Fig. \ref{FigIG}, the pump
power $P_{\mathrm{p}}$ at which this bifurcation occurs depends on the pump
frequency $\omega_{\mathrm{p}}$. This bifurcation represents the transition
between the regions \textquotesingle CC\textquotesingle $\hspace{0.025 cm}$
and \textquotesingle CR\textquotesingle $\hspace{0.025 cm}$ in the stability
map shown in Fig. \ref{FigDSM}. A bifurcation from the bistable to the
monostable regions occurs at a higher value of the pump power $P_{\mathrm{p}}%
$. This bifurcation gives rise to the sudden change in the measured response
shown in Fig. \ref{FigIG}. In the stability map shown in Fig. \ref{FigDSM},
this bifurcation corresponds to the transition between the regions
\textquotesingle CR\textquotesingle $\hspace{0.025 cm}$ and \textquotesingle C\textquotesingle.

\section{Conclusion}

We present two nonlinear effects that can be used for signal sensing and
amplification. The first one is based on the so-called
Landau-Zener-Stuckelberg process \cite{Berns_150502} of frequency mixing
between transverse and longitudinal driving tones that are simultaneously
applied to the magnon resonator. This process can be employed for frequency
conversion between the RF and the MW bands. The second nonlinear effect, which
originates from magnetization anisotropy, can be exploited for developing
intermodulation receivers in the MW band. Measurements of the intermodulation
response near the onset of the Duffing-Kerr bistability reveal a bifurcation
between a stable spiral attractor and a stable node attractor. Above this
bifurcation, i.e. where the attractor becomes a stable node, the technique of
noise squeezing can be employed in order to enhance the signal to noise ratio
\cite{Yurke_5054}.

\section{Acknowledgments}

We thank Amir Capua for helpful discussions. This work was supported by the
Russell Berrie Nanotechnology Institute and the Israel Science Foundation.

\bibliographystyle{ieeepes}
\bibliography{acompat,Eyal_Bib}

\end{document}